\documentclass[prd,amsmath,amssymb,twocolumn]{revtex4-1}
\usepackage{graphicx}% Include figure files
\usepackage{dcolumn}% Align table columns on decimal point
\usepackage{bm}% bold math
\usepackage{amssymb}% for math
\usepackage{color}

\begin{document}

\title{An exposition on Friedmann Cosmology with Negative Energy Densities}

\author{Robert J. Nemiroff}
%\email{nemiroff@mtu.edu}
\affiliation{Department of Physics, Michigan Technological University, 1400 Townsend Drive, Houghton, MI 49931}
\author{Ravi Joshi}
%\email{rjoshi@mtu.edu}
\affiliation{Department of Physics, Michigan Technological University, 1400 Townsend Drive, Houghton, MI 49931}
\author{Bijunath R. Patla}
\email{bijunath.patla@nist.gov}
\affiliation{National Institute of Standards and Technology, 325 Broadway, Boulder, CO 80305}

\begin{abstract}
%{\color{red} say like this}
 How would negative energy density affect a classic Friedmann cosmology?  Although never measured and possibly unphysical, certain realizations of quantum field theories leaves the door open for such a possibility. In this paper we analyze the evolution of a universe comprising varying amounts of negative energy forms.
 % ($\Omega < 0$). 
Negative energy components have negative normalized energy densities, $\Omega < 0$. They include negative phantom energy with an equation of state parameter $w<-1$, negative cosmological constant: $w=-1$, negative domain walls: $w=-2/3$, negative cosmic strings: $w=-1/3$, negative mass: $w=0$, negative radiation: $w=1/3$ and negative ultralight: $w > 1/3$. 
 Assuming that such energy forms generate pressure like perfect fluids, the attractive or repulsive nature of negative energy components are reviewed. The Friedmann equation is satisfied only when negative energy forms are coupled to a greater magnitude of positive energy forms or positive curvature.
 %; minimal cases of both of these yield analytic solutions involving hypergeometric functions. 
 We show that the solutions exhibit cyclic evolution with bounces and turnovers.The future and fate of such universes in terms of curvature, temperature, acceleration, and energy density are reviewed. 
The end states are dubbed ``big crunch," `` big void," or ``big rip" and further qualified as ``warped",``curved", or ``flat",``hot" versus ``cold", ``accelerating" versus ``decelerating" versus ``coasting". A universe that ends by contracting to zero energy density is termed "big poof." Which contracting universes ``bounce" in expansion and which expanding universes ``turnover" into contraction are also reviewed.
\end{abstract}

\begin{description}
%\item[To be submitted to] \prd
\item[PACS numbers]{98.80.-k,98.80.Bp,98.80.Cq,98.80.Qc}
\end{description}

\maketitle

\section{The Friedmann Equation of Energy, Expanded to Negative Energies}

The classic Friedmann equation of energy  is typically written in the form \cite{weinberg72, peebles88,peacock99,mukhanov05}
\begin{equation} \label{FriedmannOld}
  H^2 = {8 \pi G \over 3} \rho - {k\over R^2} ,
\end{equation}
where $H$ is the Hubble parameter, $G$ is the gravitational constant, $\rho$ is the energy density, $R$ is a scale factor of the universe, and $k$ is a dimensionless constant related to the curvature of the universe. The speed of light $c$ is set to unity. The Hubble parameter $H = {\dot R} / R = {\dot a / a}$, where $a$ is the dimensionless scale factor of the universe such that $a = R / R_0$ and $R_0$ is the scale factor of the universe at some canonical time $t_0$.

In \cite{nemiroff08} (hereafter Paper I), the average energy density $\rho$ was explicitly expanded into all possible stable {\it positive} energy forms, including phantom energy, cosmological constant, domain walls, cosmic strings, compact matter, radiation, and hypothetical energy forms collectively dubbed ultralight. These ``stable" energy forms are considered different than energy fields such as scalar fields in that each has a well defined local form and evolves in the universe as an constant power of $a$. All energy forms evolve as an integer power of the scale factor $a$.

In this paper, we consider stable energy forms with {\it negative} energy density, $\rho < 0$. To the best of our knowledge, some of these energy forms have never been discussed explicitly. It is not being suggested that any of these energy forms are presently important in the universe, but their  theoretical inclusion presents cosmologically interesting scenarios. A universe comprising mostly of negative energy density component will perhaps violate various energy conditions of general relativity, and may even violate the second law of thermodynamics~\cite{hawking}. But some extensions of quantum field theory and recent studies of Cassimir force experiments indicate  that energy conditions in relativity need only be satisfied on a global scale, or on an average measure, leaving the possibility of violation of the very same energy conditions locally or for a small period of time~\cite{epstein65,roman,ford96,pfenning,helfer,visser00, graham,fewster}. 
Therefore, for  example, if some physical process that might have lasted only for a very short time compared to the age of the physical universe and may also have violated relativistic energy conditions could still play a role in the evolution of the universe by virtue of producing small amounts of negative energy components in the past that could become relevant at a later time in the evolution of such universes.

Following Paper I, the total energy density $\rho$ can be expanded
into its component stable forms such that
\begin{equation} \label{rho}
  \rho = \sum_{n=-\infty}^{\infty} \rho_n^+ \, a^{-n}
              + \sum_{m=-\infty}^{\infty} \rho_m^- \, a^{-m},
\end{equation}
where $\rho_n^+$ is the usual positive energy density typically considered and $\rho_m^-$ is the negative cosmological energy density. In general, in this work, the subscript $n$ and the (sometimes redundant superscript plus sign) will refer to a positive energy density, while the subscript $m$ (and superscript minus sign) will refer to a negative energy density. Note that the values of $\rho_n^+$ and $\rho_m^-$ remain fixed to their values at $a = 1$ and do not change as the universe evolves---hence their designation as ``stable".

As usual, a normalizing (positive) critical density $\rho_c$ is defined such that $\rho_c = 3 H^2/(8 \pi G)$ so that $\Omega = \rho/\rho_c$. Dividing both sides of Eq.~(\ref{rho}) by $\rho_c$ at $a=1$, we obtain
\begin{equation} \label{OmegaSum}
\Omega  = \sum_{n=-\infty}^{\infty} \Omega_n^+ \, a^{-n} +  \sum_{m=-\infty}^{\infty} \Omega_m^- \, a^{-m} .
\end{equation}
Following above convention, also used in Paper I, $\Omega$ depends on $a$ and hence the time $t$. 
%The values of $\Omega_n^+$ and $\Omega_m^-$ remain fixed at the time when $a=1$. 
Note that $\Omega$ is different from $\Omega_{total}$, the sum of all of the stable $\Omega_n^+$ and $\Omega_m^-$ values, where
\begin{equation}
\Omega_{total} = \sum_{n=-\infty}^{\infty} \Omega_n^+  +
\sum_{m=-\infty}^{\infty} \Omega_m^-  .
\end{equation}

The curvature term in Eq. (\ref{FriedmannOld}) can be written in terms of more familiar quantities. We divide Eq. (\ref{FriedmannOld}) by $H^2$ to obtain 
\begin{equation}
{ k c^2 \over H^2 R^2 }  = \Omega - 1 .
\end{equation}

Given this nomenclature, a more generalized Friedmann equation of energy can be written in a dimensionless form that explicitly incorporates all possible stable, static, isotropic energy forms described by an integer $n$. 
Using $H={\dot a}/a$ in Equation~(\ref{FriedmannOld}) and dividing each side by the square of the Hubble parameter $H_0^2$ (at $a=1$), we obtain
\begin{eqnarray} \label{friedmanna}
\Big(\frac{\dot a}{H_0}\Big)^2 =
(1 - \Omega) +\sum_{n,m=-\infty}^{\infty} (\Omega_n^+ \, a^{2-n}
                  + \Omega_m^- \, a^{2-m}) .
\end{eqnarray}
Note that Eq.(\ref{friedmanna}) can be written to highlight only present day observables by substituting $1/a = R_0/R = (1+z)/(1+z_0) = (1+z)$, where $z$ is the redshift and $z_0 = 0$ is the redshift at the epoch where $a=1$.

\section{Description of negative energy components based on pressure and equation of state parameter}

%As reviewed in Paper I, energy is not conserved in general relativity on the global scale. For satisfying conservation of energy locally, it is often convenient to choose a frame that is at rest with respect to the isotropy of cosmological energies. As an example, for radiation, this rest frame at any point in the universe is where the cosmic microwave background radiation (CMB) appears uniform  and isotropic in intensity and temperature.

%\subsection{Perfect Fluids and Gravitational Pressure}

The Friedmann equations typically quantify the cosmological evolution of perfect fluids. Perfect fluids are characterized by only two variables: energy density $\rho$ and isotropic pressure $P$. These two variables can be isolated in an effective modified Poisson equation for gravity from general relativity in the weak field limit so that\cite{weinberg72,peacock99}
\begin{equation} 
\label{poisson}
\nabla^2 \phi = \frac{4 \pi G }{c^2}(\rho + 3 P) ,
\end{equation}
where $\phi$ is the Newtonian potential in the limit of weak field gravity. The gravitational influence of pressure has no Newtonian analogue.

In general, the equation of state of a perfect fluid is given by $w = P/\rho $. Local conservation of the energy-momentum tensor  implies that $\rho \propto a^{-3(1+w)}$, where $n= 3(w + 1)$ and $w = n/3 - 1$. Locally conserved perfect fluids will be assumed by default for the rest of this paper.

Positive energy densities with $w<0$ ($n<3$), typically referred to as forms of ``dark energy", have formally negative gravitational pressure which corresponds to repulsion of a like energy. For positive points of energy when $n = 3$ and so $w = 0$, there is no gravitational pressure. Positive energy densities with $w > 0$ ($n > 3$), here referred to as ``light energy", have formally positive gravitational pressure which corresponds to an attraction of a like energy.

Conversely, {\it negative} energy densities that have $w<0$ ($n<3$), here also referred to as forms of ``dark energy", have formally {\it positive} gravitational pressure which corresponds to attraction of a like energy. For negative points of energy when $n = 3$ and so $w = 0$, there is no gravitational pressure. Negative energy densities with $w > 0$ ($n > 3$), also here referred to as ``light energy", have formally negative gravitational pressure which corresponds to a repulsion of a like energy.

%\subsection{Minimally simplified Friedmann Equations and Solutions}

%In general, both Friedmann equations will be codified into a single line denoted the ``Friedmann equation", referring predominantly to the Friedmann equation of energy which was shown above as Eq.(\ref{friedmanna}).

Two specific cases will be considered here, the first of which will be that with zero curvature (``flat"), where a single stable negative energy form is coupled with a positive energy form. The Friedmann equation in this case is
\begin{equation} 
\label{friedmannflat}
{ {\dot a}^2 \over H_o^2 } =
\Omega_n^+ a^{2-n} + \Omega_m^- a^{2-m} ,
\end{equation}
where $\Omega_n^+>0$ and $\Omega_m^-<0$ are the normalized values of positive and negative energy densities. The flatness condition demands that $\Omega_n + \Omega_m = 1$. The solution to Eq.(\ref{friedmannflat}) is (for a more general discussion, see \cite{lake06})
\begin{equation}
 t - t_o = {1 \over H_o} \int_{a_o}^a { da \over
         \sqrt{ \Omega_n^+ a^{2 - n} + \Omega_m^- a^{2 - m} }  },
\end{equation}
where $a_o$ is the value of the  scale factor at time $t_o$ and $\Omega_n a^{2-n} + \Omega_m a^{2-m}>0$.

The second  case considered here will be a single stable type of negative energy alone in the universe coupled only with the positive curvature it creates. The Friedmann equation in this case is
\begin{equation} \label{friedmanncurved}
{ {\dot a}^2 \over H_o^2} =
(1 - \Omega_m) + \Omega_m a^{2-m} ,
\end{equation}
where by definition $\Omega_m < 0$. Written in integral form
\begin{equation}
 t - t_o = {1 \over H_o} \int_{a_o}^a { da \over
             \sqrt{ (1 - \Omega_m) + \Omega_m a^{2-m} } } ,
\end{equation}
where $(1 - \Omega_m) + \Omega_m a^{2-m} >0$.

%In the subsequent sections specific stable energy forms involving different integer values of $n$ and $m$ will be discussed. Analytic solutions exist for single component
% Put a limit on the magnitude of different current negative
% energy density terms from observations in modern cosmology.

\section{Friedmann Equation for Stable Negative Energy Forms}
\label{sec_fried}

Energy can take any number of stable forms, and is usually known by what form it takes. With respect to the Friedmann Equations, forms of energy can be classified by how they affect the universe gravitationally. For example, energy forms represented by integer values of $n$ and corresponding equation of state parameter $w$ for positive energy density is given in Table~\ref{tab1a}.
\begin{table}[ht]
 \begin{center}
  \begin{tabular}{  l l  l l}
  \hline\hline\\
  [-1.0ex]
  $w$\hspace{1cm} & $n$\hspace{0.5cm} & Dimensional Type\hspace{0.3cm} & Name \\[1.0ex]
  \hline\\[-1.5ex]
  $<-1$ & $<0$ & Unknown &Phantom Energy \\
  %  \hline
  $-1$ & $0$ & Volume &Cosmological Constant\\
%  \hline
  $-2/3$ & $1$& Sheet &Domain Wall \\
% \hline
  $-1/3$ & $2$& Line &Cosmic String \\
%  \hline
  $0$ & $3$&Point &Matter\\ 
%  \hline
  $1/3$ & $4$&Relativistic Point &Radiation\\
%  \hline
  $>1/3$ & $>4$&Unknown &Ultralight\\\\[-1.5ex]
  \hline
 % \\[1.0ex]
    \end{tabular}
	\caption  {\label{tab1a}Energy forms that may drive the expansion rate of the universe as described by the Friedmann Equations.The entries include known and hypothesized energy forms that are thought to be stable perfect fluids over cosmological scales}
 \end{center}
\end{table}
It is assumed that all stable energy forms remain stable with a density directly extrapolated from the density at $a = 1$. Although $a=1$ could describe any epoch, here it will typically describe the present. Many energy terms used below have precedents and are in common use.

Although no closed form solutions exist for Eq.(\ref{friedmanna}) in general, analytical solutions involving hypergeometric functions exist for particular cases:
 a) single energy type corresponding to positive and negative energy densities existing in a flat universe ($\Omega_n^+ +\Omega_m^- =1$)  and b) a single negative energy type in a curved universe ($\Omega_m^- -1\ne0$). 
 
 \subsection{Solutions for a single positive and negative energy component in a flat universe}
 A universe that consists of only a positive and a negative energy density type satisfies
Eq.(\ref{friedmannflat}), with the solution
%\begin{widetext}
\begin{equation}
\label{hypgeo}
H_0t(a)=\frac{2}{n}\left(\frac{a^n}{\Omega_n^+}\right)^{1/2}\,_2F_1\left(1/2,p,p+1,q\right)
\end{equation}
%\end{widetext}
where
\begin{equation}
p\equiv \frac{n}{2(n-m)}, \quad q \equiv  -\frac{\Omega_m^-}{\Omega_n^+}a^{n-m} \ge 0.
\end{equation}
Generally, the validity of the solution corresponds to values of $p\ge0$. Negative values of $p$ are allowed, but the hypergeometric function has a series of line singularities along the negative axis. The physical interpretation of singularities point toward cyclic universes resulting from a bounce following a period of contraction or 2) turnaround following an expansion.
\begin{equation}
p>0 \implies \begin{cases}   n>m,   & n>0 \\
  n<m, & n<0.
 \end{cases}
 \label{p_cond}
 \end{equation}
$p=0$ when $n=0$, and when $n=m$, $p\rightarrow\infty$.

\begin{table}[ht]
 \begin{center}
  \begin{tabular}{  l l  c }
  \hline\hline\\
  [-1.0ex]
  $p$\hspace{1.5cm} & $q$\hspace{1.0cm} & $_2F_1(1/2,p,p+1,q)$ \\[1.0ex]
  \hline\\[-1.5ex]
  $\infty$ 				& $\infty$ 		& 0 \\
%  $p<-1$               &  $\infty$     & Indeterminate\\
   $\infty$ 				& 0        		& 1 \\
  0        				& 0        		& 1 \\
  0        				& $\infty$ 		& 1 \\
 $p\rightarrow -1 $      & 0             & 1 \\ 
 $p\rightarrow -1 $      & $\infty$      & $\infty$ \\  
  \hline
 % \\[1.0ex]
    \end{tabular}
	\caption  {\label{tab2a}The value of $p$ as defined in Eq.(\ref{hypgeo}) determines the allowed combination of positive and negative energy types. The hypergeometric function has a singularity at $p=-1$. The values of $q$ depend on the scale factor in an expanding or contracting universe. $_2F_1$ has a series of singularities after $p\le-1$. }
 \end{center}
\end{table}

As a particular case, for $|q|<1$, that is, when $a\sim 1$ and $|\Omega_m^-|< |\Omega_n^+|$
\begin{equation}
_2F_1\left(1/2,p,p+1,q\right)=\sum_0^\infty \frac{(1/2)_j (p)_j}{(p+1)_j}\frac{q^j}{j!}
\label{hyp-geo}
\end{equation}
where $_2F_1$ is the hypergeometric function (with also a series expansion interms of Legendre function\cite{erdelyi}) and 
\begin{equation}
(x)_j = \begin{cases}   1   & j = 0 \\
  \prod_{i=1}^{j}(x+j-1) & j > 0.
 \end{cases}
 \end{equation}
Ignoring higher order terms in q, Eq.(\ref{hypgeo}) takes the form
\begin{equation}
t=H_0^{-1}\frac{2}{n}\left(\frac{a^n}{\Omega_n^+}\right)^{1/2}\left(1+\frac{\Omega_m^-}{\Omega_n^+}\left(\frac{n}{2}\right)\frac{a^{n-m}}{3n-2m}\right).
\end{equation} 
%We note that for the case $m=n$ the solution yields low time scales, therefore, in terms of stability of the solutions it is not in contradiction with Eq.(\ref{p_cond}). The values for single component universe with and without bounce are identified in Table~\ref{tab3a}. $_2F_1$ as a function of $p$ is shown in fugure~\ref{fig:1}.
Also, see~\cite{spokoiny93,steinhardt02,dabrowski06,cai10} for reviews on oscillating solutions describing some cyclic models. 
\begin{figure}[htb]
%\begin{center}
%\includegraphics{hypgeo.eps}
\includegraphics[width=0.5\textwidth]{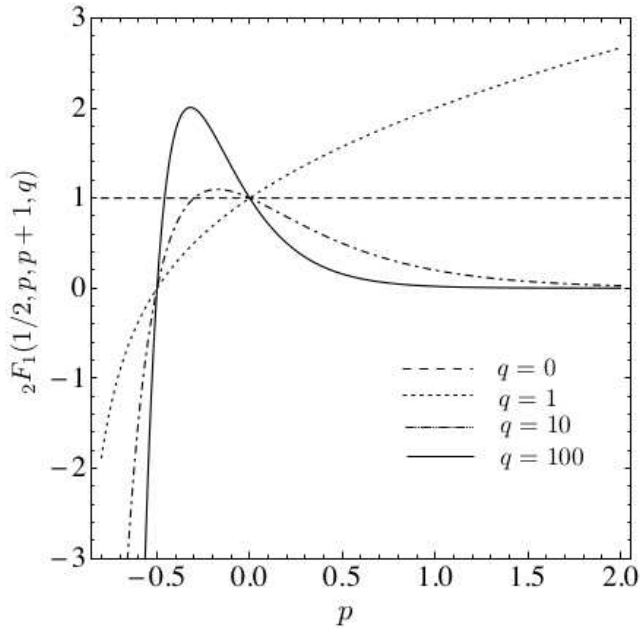}
\caption{The value of the hypergeometric function has a series of singularities along the negative x-axis for values of $p$, $q$ is a function of $m$ and the scale factor $a$. Note that if $a=0$ then there are no singularities and the function is independent of $m$ and $n$. This suggests that the universes that bounce or turns around does so before $a$ becomes zero. }
\label{fig:1}
%\end{center}
\end{figure}

\subsection{Solution for a negative energy component in a curved universe}
The solution for the Friedmann equation involving a negative energy density component coupled to curvature, Eq.(\ref{friedmanncurved}), is

\begin{equation}
\label{hypgeo_curved}
H_0t(a)=\left(\frac{a^2}{1-\Omega_m^{-}}\right)^{1/2}\,_2F_1\left(1/2,p_{_k},p_{_k}+1,q_{_k}\right)
\end{equation}
where
\begin{equation}
p_{_k}\equiv \frac{1}{2-m}, \quad q_{_k} \equiv  -\frac{a^{2-m}}{1-\Omega_m} \le 0.
\end{equation}
The solutions in this category without a bounce correspond to values of $m<3$, which are  -1, 0, 1 and  2. 
%Therefore, only negative phantom energy, negative cosmological constant and negative cosmic strings can couple with curvature to form a stable universe.
We note that only numerical solutions exist for many component curved universe given by Eq.(\ref{friedmanna}).

\subsection{Possible Futures for Universes}

An evolving Friedmann universe may be described in terms of its ultimate fate, curvature, temperature, and energy density as the scale factor $a$ approaches its maximal value, by fully analyzing Eq.(\ref{friedmanna}).

First, the fate of each universe will be analyzed. In other words, what becomes of scale factor $a$? An expanding universe can either expand forever, meaning that scale factor $a$ going to infinity is formally allowed by the Friedmann equations, or  ``turnover", meaning that a maximum $a_{max}$ will be reached, after which the universe will collapse. Similarly, a collapsing universe can contract to a point, meaning that scale factor $a$ going to zero is formally allowed by the Friedmann equations, or ``bounce", meaning that a minimum $a_{min}$ will be reached, after which the universe will expand.

%Next, the curvature of the universe will be characterized. In other words, what becomes of curvature? If the curvature term $(1 - \Omega_n^+ - \Omega_m^-)$ (of Eq. (\ref{friedmanna}), for example), the term that is directly added to the Friedmann energy terms, goes to zero as $a$ goes to either zero or infinity, even if predetermined by a flat initial condition to always be zero, then this universe will be deemed to have a ``Flat" ending. Conversely, if this curvature term diverges as $a$ goes to either zero or infinity, then this universe will be deemed to have a ``Warped" ending. If the curvature term ends up at a finite value, then the universe will be deemed to have a ``Curved" ending.

Next, the temperature of the universe will be characterized. Here temperature will be considered directly related to energy density. If the positive energy density term $\Omega_n a^{-n}$ diverges as $a$ goes to either zero or infinity, then this universe will be deemed to have a ``hot" ending. Conversely if this positive energy density term goes to zero as $a$ goes to either zero or infinity, then this universe will be deemed to have a ``cold" ending. If the positive energy density ends up at a finite value, then this universe will be deemed to have a ``warm" ending.

Next, the kinematics of the universe will be characterized. In other words, what happens to the expansion speed ${\dot a}$ of a universe? 
%To determine this, the Friedman Equation will be multiplied by $H_o^2 a^2$ to isolate the ${\dot a}^2$ term so that
%\begin{equation}
%{\dot a}^2 = H_o^2 [(1 - \Omega_n - \Omega_m) a^2
% + \sum_{n=-\infty}^{\infty} \Omega_n^+ a^{2-n}
% + \sum_{m=-\infty}^{\infty} \Omega_m^- a^{2-m}] .
%\end{equation}
If ${\dot a}^2$ diverges as $a$ goes to either zero or infinity, then this universe will be deemed to have an ``accelerating" ending. If ${\dot a}^2$ drops to zero as $a$ goes to either zero or infinity, then this universe will be deemed to have a ``decelerating" ending. If ${\dot a}^2$ ends with a constant value, then this universe will be deemed to be ``coasting" as its end.

Alternatively,  if the universe ends with $a$ going to infinity but the energy density decreasing to zero then this universe ending will be deemed a ``big void". This term is here preferred over the previously popular ``big freeze" because it is more descriptive of the end state, as the end temperature of the universe has already been characterized. If $a$ goes to infinity but the energy density remains constant, then this universe ending will be deemed to result in a ``steady state". If a universe ends with $a$ going to infinity but the energy density also diverges, possibly in a finite time, this universe ending will be deemed a ``big rip".

The end states of contracting universes will also be classified. If the universe ends with $a$ going to zero but the energy density diverges, this universe ending will be deemed a ``big crunch". If the universe ends with $a$ going to zero but the energy density remains constant, then this universe will (again) approach a ``Steady State". And finally if the universe ends with $a$ going to zero and the energy density also drops to zero, this universe ending will be deemed a ``big poof". To the best of our knowledge, ``big poof" universes have not been discussed previously. The criteria adopted to describe the future of universes that are expanding or contracting is summarized in Table~\ref{tab3a}.

\begin{table*}[ht]
 \begin{center}
  \begin{tabular}{  l| l l l l |  l l l}
  \hline\hline\\
  [-1.0ex]
  Universe\hspace{1.5cm} & $n$\hspace{1.0cm} & $\Omega_n a^{-n}$\hspace{1.0cm} & End State\hspace{1.0cm} & Temperature\hspace{1.0cm}  &$n$\hspace{1.0cm} & $\dot{a}\propto\Omega_n a^{2-n}$\hspace{0.5cm} & Kinematics\\[1.0ex]
  \hline\\[-1.5ex]
  Expanding              & $n>0$ & $\sim 0$      & Big Void         & Cold   & $n>2$ & $\sim 0$       & Decelerating\\
  Expanding              & $n<0$ & $\sim \infty$ & Big Rip          & Hot    & $n<2$ & $\sim \infty$  & Accelerating\\
  Expand/Contract        & $n=0$ & constant      & Steady State     & Warm   & $n=2$ & constant       & Coasting\\
  Contracting            & $n>0$ & $\sim \infty$ & Big Crunch       & Hot    & $n>2$ & $\sim \infty$  & Accelerating \\
  Contracting            & $n<0$ & $\sim 0$      & Big Poof         & Cold   & $n<2$ & $\sim 0$       & Decelerating\\         
             \hline
 % \\[1.0ex]
    \end{tabular}
	\caption  {\label{tab3a}Possible fates of universes---characterized by "appropriate" description for end state and temperature based on normalized energy density ($\Omega_n a^{-n}$), and kinematics based on the rate of scale factor ($\Omega_n a^{2-n}$).}
 \end{center}
\end{table*}

%\subsection{Possible Pasts for Universes}

%The pasts of Friedmann universes including negative energy universes will also be characterized in terms of beginning state, curvature, temperature, and energy density when the scale parameter $a$ starts at its minimal value. In similarity to the above-discussed futures of universes, a universe can have either a flat, curved, or warped beginning, either a cold, warm, or hot beginning, and either an accelerating, coasting, or decelerating beginning.

%As with the discussion of universe futures, we will focus on the initial state of the energy density and the scale factor of the universe to describe its initial character. 

To describe the initial condition of a universe, the term ``big bang" will be extended to all universes that start with $a=0$ even if the energy density started with a finite or formally zero value. However, were the universe to start from a finite value of $a$, which might have followed a contraction phase, this universe will be deemed to have started from a ``bounce."

\begin{figure}[t]
%\begin{center}
%\includegraphics{lcdm.eps}
\centering
\includegraphics[width=0.5\textwidth,scale=0.5]{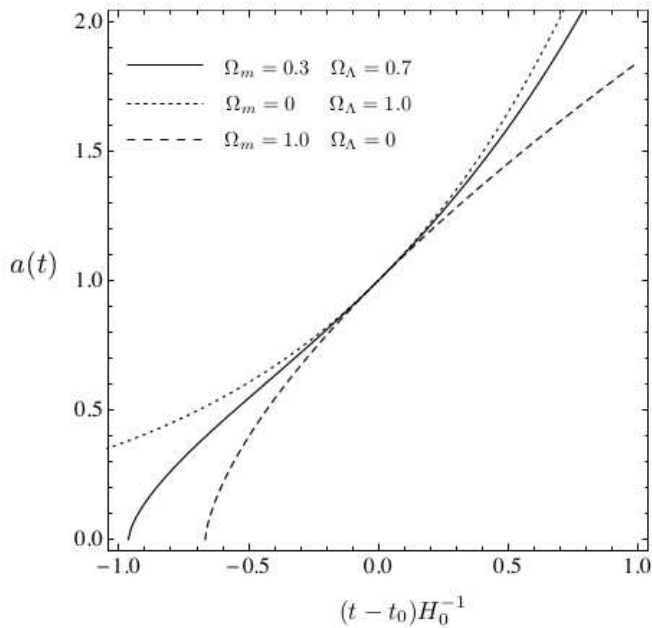}
\caption{As a prelude to providing numerical solutions to Eq.~(\ref{friedmanna}), wherever appropriate, we verify it with the known energy composition of the observable universe. Note that it gives a very familiar age for the universe corresponding to the value of zero scale factor. A  de-Sitter universe and a matter dominated universe is given for comparison.}
\label{fig:2}
%\end{center}
\end{figure}

Negative energy component alone will not satisfy the Friedmann equation---resulting in purely imaginary solutions. Using the formalism adopted in this paper and also to validate Eq.~(\ref{friedmanna}) using the known energy composition of the $\Lambda$CDM universe, we provide known solutions in figure~\ref{fig:2}.

In the following section, we analyze negative energy components coupled with   positive energy components in a flat or curved universe.  As we have shown in section~\ref{sec_fried}, single component flat universes comprising a single negative and positive energy density component, and  curved universe coupled to a negative energy density component have analytical solutions.  We discuss asymptotic behavior and as well as the nature of the solutions near critical points that correspond to values of scale factors that are a local extremum for these solutions. 

\section{Asymptotic behavior of universes containing varying fractions of positive and negative energy densities}

The friedmann equation with both positive and negative energy density terms in a flat universe will attain a critical value of scale factor
\begin{equation}
a^{\rm{flat}}_{crit} = \left(\frac{\Omega_{m}^- }{\Omega_{m}^- -1}\right)^{1/(m-n)},
\label{crit_sf_flat}
\end{equation}
where $m$ is an index corresponding to negative energy component. The corresponding equation for a curvature dominated universe is 
\begin{equation}
a^{\rm{curved}}_{crit} = \left(\frac{\Omega_{m}^- -1 }{\Omega_{m}^- }\right)^{1/(2-m)}.
\end{equation}
The end states corresponding to different combinations of $n$ and $m$ for expanding and contracting universes,  for curvature coupled negative energy density universes and for single component universes are summarized in Tables~\ref{tab7a} and, \ref{tab5a} and  \ref{tab6a}   respectively. 
Below we present a brief discussion on the properties of some of these negative energy forms.

\begin{table}[htb]
 \begin{center}
\begin{tabular}{lll}
	\hline
	\hline\\
	[-1.5ex]%	\vspace{-0.13in}
Universe  \hspace{0.4in} &$m$\hspace{0.4in} & description \\
	[1.5ex]	
	\hline\\
Expanding   & $m<2$ & turns over\\
			&  $m \ge 2$ & Cold Coasting Big Poof\\
Contracting & $m\le 2$   & Hot Coasting Big Crunch\\
			& $m>2$ & bounce\\			
			
	\hline
	\end{tabular}
		\caption{\label{tab7a}How do two stable energy forms interact to influence the fate of a curved universe. }
%		\end{tabular}
 \end{center}
\end{table}

\begin{table*}[h]
 \begin{center}
	\begin{tabular}{ c  p{2.2cm} p{2.2cm} p{2.2cm} p{2.2cm} p{2.2cm} p{2.2cm} p{2.2cm} p{2.2cm}}
	\hline
	\hline\\
	[-1.5ex]%	\vspace{-0.13in}
%{$m$}{$n$} 
\raisebox{-5pt}{$m$}\hspace{10pt}\llap{\raisebox{10pt}{$n$}} & -1 & 0 & 1& 2 & 3 & 4 & 5 \\
%	m &    &   &  &   &   &   &\\
	[1.5ex]	
	\hline\\
	-1 &Hot Accelerating Big Rip
 	& turns over & turns over & turns over & turns over & turns over & turns over \\
	\hline\\
	0 & Hot Accelerating Big Rip & Warm Accelerating Steady State & turns over & turns over & turns over & turns over & turns over\\
	\hline\\
	1 &Hot Accelerating Big Rip & Warm Accelerating Steady State &Cold Accelerating Big Void.
 	& turns over & turns over & turns over & turns over \\
	\hline\\
	2 &Hot Accelerating Big Rip & Warm Accelerating Steady State &Cold Accelerating Big Void &Cold Coasting Big Void.
	 & turns over & turns over & turns over \\
	\hline\\
	3 &Hot Accelerating Big Rip & Warm Accelerating Steady State &Cold Accelerating Big Void &Cold Coasting Big Void &Cold 	Decelerating Big Void & turns over & turns over \\
	\hline\\
	4 &Hot Accelerating Big Rip & Warm Accelerating Steady State &Cold Accelerating Big Void &Cold Coasting Big Void &Cold Decelerating Big Void &Cold Decelerating Big Void &turns over \\
	\hline\\
	5 &Hot Accelerating Big Rip & Warm Accelerating Steady State & Cold Accelerating Big Void &Cold Coasting Big Void &Cold Decelerating Big Void &Cold Decelerating Big Void &Cold Decelerating Big Void \\
	\hline
	\end{tabular}
		\caption{\label{tab5a}How do two stable energy forms interact to influence the fate of a flat  expanding Universe. }
 \end{center}
\end{table*}

\begin{table*}[h]
 \begin{center}
\begin{tabular}{ c  p{2.2cm} p{2.2cm} p{2.2cm} p{2.2cm} p{2.2cm} p{2.2cm} p{2.2cm} p{2.2cm}}
	\hline
	\hline\\
	[-1.5ex]%	\vspace{-0.13in}
%{$m$}{$n$} 
\raisebox{-5pt}{$m$}\hspace{10pt}\llap{\raisebox{10pt}{$n$}} & -1 & 0 & 1& 2 & 3 & 4 & 5 \\
%	m &    &   &  &   &   &   &\\
	[1.5ex]	
	\hline\\
	-1 &Cold Decelerating Big Poof & Warm Decelerating Steady State & Hot Decelerating Big Crunch & Hot Coasting Big Crunch & Hot Accelerating Big Crunch & Hot Accelerating Big Crunch & Hot Accelerating Big Crunch \\
	\hline\\
	0 & bounce & Warm Decelerating Steady State & Hot Decelerating Big Crunch & Hot Coasting Big Crunch & Hot Accelerating Big Crunch & Hot Accelerating Big Crunch & Hot Accelerating Big Crunch \\
	\hline\\
	1 &bounce & bounce & Hot Decelerating Big Crunch & Hot Coasting Big Crunch & Hot Accelerating Big Crunch & Hot Accelerating Big Crunch & Hot Accelerating Big Crunch \\
		\hline\\
	2 & bounce & bounce & bounce & Hot Coasting Big Crunch & Hot Accelerating Big Crunch & Hot Accelerating Big Crunch & Hot Accelerating Big Crunch \\
	\hline\\
	3 & bounce & bounce & bounce & bounce & Hot Accelerating Big Crunch & Hot Accelerating Big Crunch & Hot Accelerating Big Crunch \\
	\hline\\
	4 & bounce & bounce & bounce & bounce & bounce &Hot Accelerating Big Crunch & Hot Accelerating Big Crunch \\
	\hline\\
	5 & bounce & bounce & bounce & bounce & bounce & bounce &Hot Accelerating Big Crunch \\
	\hline
	\end{tabular}
		\caption{\label{tab6a}How do two stable energy forms interact to influence the fate of a flat contracting universe. }
%		\end{tabular}
 \end{center}
\end{table*}

The existence of any form of phantom energy ($n < 0$ or $m < 0$, and hence $w < -1$) is controversial as it may violate general relativistic energy conditions \cite{cladwell03,nemiroff08}.
Negative phantom energy would have $\rho < 0$ so that the gravitational energy density would be repulsive. However, since $w < -1$, this positive energy density is necessarily coupled to a gravitational pressure that is attractive and, from Eq.(\ref{poisson}), three times stronger. Therefore, on the whole, phantom energy would gravitationally attract itself.

When $n = m = -1$, since $\Omega_n + \Omega_m = 1$, if a contracting universe remains dominated by the repulsive positive phantom energy, net positive phantom energy density decreases as the scale factor $a$ decreases. Note that this can occur  as a time-reversed version of  $n = m = -1$ expanding universe ending in a flat hot accelerating big rip.
% is just a milestone and does not designate an end time for a universe evolving with time somehow greater that $t_{Rip}$ discussed in Eq. (\ref{phantom_time}). 
Either way, such a contracting universe will asymptotically approach zero scale factor and zero energy density over an infinite time interval. This universe will end in a flat cold decelerating ``big poof" ---an unrealistic scenario in our universe, but nevertheless a possibility in at least some (out of tens of millions) universes in the (presently accepted) inflationary paradigm\cite{guth81,spergel03}.

A negative cosmological constant has $m = 0$ ($w = -1$) energy with gravitationally attractive pressure. Although an $m = 0$ component has a negative energy density that is gravitationally repulsive, Eq.(\ref{poisson}) shows that the pressure is positive and three times greater in magnitude. Therefore the net effect of negative stable cosmological constant energy on itself is attractive. 

A test particle(point mass) near a static cosmological domain wall of negative energy would experience a gravitational attraction toward it. This is the opposite of the gravitational repulsion a person would feel toward a positive energy domain wall\cite{davis87,turok89,kamionkowski90,preskill92}.

A negative cosmic string has $m = 2$ ($w = - 1/3$) energy with gravitationally attractive pressure. It is conceivable that the centers of negative energy cosmic strings contains wormholes with boundary conditions satisfying Einstein's equation\cite{aros97}. Although the density of  $m = 2$ energy is negative, and is gravitationally repulsive, Eq.(\ref{poisson}) shows that the attractive pressure is equal in magnitude. If a person (point particle) were to go up to a flat, static, cosmic string of negative energy, that person would feel neither gravitational attraction toward it nor repulsion away from it. This is similar to the effect a person would feel toward a positive energy cosmic string\cite{vilenkin81,vilenkin81-2}.

As discussed in Paper I in more detail than given here, a fundamental type of energy which evolves as $n = 3$ ($w = 0$)  has energy confined to regions small compared to the gravitational horizon size\cite{riess05}. These points do not change in energy as the universe either expands or contracts, as fundamental particles like the electron do not change their energy (mass) as the universe expands. Such stable points of energy are called matter and are fundamentally different than other forms of energy. 
A negative point mass, as indicated by $w = 0$, has no gravitational pressure. A summary of how $w = 0$ negative energy interacts with other forms of energy is given in Table 1. Were a person were to go up to a negative point mass, that person would feel gravitational repulsion from it. This is opposite of the gravitational repulsion a person would feel toward a positive point mass. Were negative point masses to be dropped near the Earth, they would accelerate toward the center of the earth due to their negative inertial mass, although the earth's gravity would exert a force that is opposite in direction to this acceleration. This is one aspect of the equivalence principle.

Stable negative energy forms with $w > 1/3$ ($n > 4$), discussed recently for stable positive energy forms was referred to as ultralight \cite{nemiroff08}. The term ultralight energy contrasts with dark energy as being beyond light in terms of gravitational pressure. By analogy, ultralight is to light what ultraviolet light is to violet light.
The idea of energy evolving with an effective $w > 1/3$ has been hypothesized previously in the context of time-varying scalar fields (see for example \cite{guth85,sahni02,steinhardt03}). However, the {\it stable} negative ultralight energy forms discussed in this section are specifically constrained in their behavior to evolve {\it only} in accordance with the Friedmann equations are not scalar fields.
Forms of ultralight with $w > 1$ might be unphysical because their formal sound $c_s = c \sqrt{w}$ is greater than the speed of light. Only ultralight energy forms with $1/3 < w < 1$ would have a sound speed less than c.

\begin{figure}[t]
%\begin{center}
%\includegraphics{phantom-neg.eps}
\includegraphics[width=0.45\textwidth]{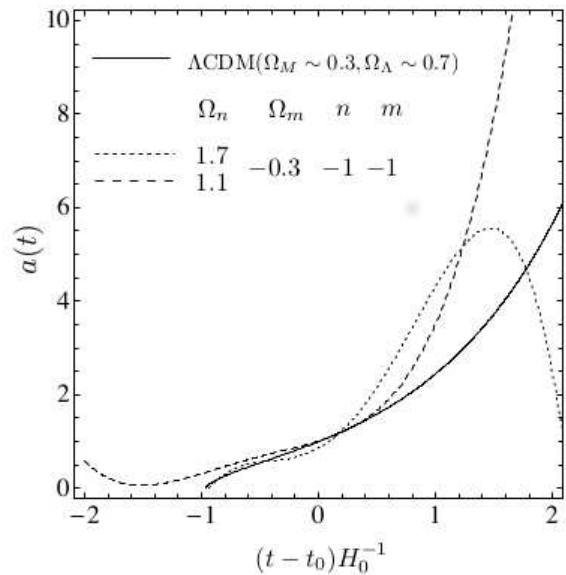}
\caption{Numerical solution: For a curved universe dominated by both positive and negative phantom energy. }
\label{fig:3}
%\end{center}
\end{figure}

As previously noted, we have to rely on numerical solutions for multicomponent (with more than on energy form corresponding to $n$ or $m$ with our without curvature) universes. As an example, numerical estimation of the scale factor for a curved universe with both positive and negative phantom energy is shown in figure~\ref{fig:3}.

\section{On the gravitational interaction between a positive and a negative Energy Species}

As an example, consider $n = 2$ and $m = 0$. Here the positive $n = 2$ energy is gravitationally neutral and will neither attract nor repel other $n = 2$ energies. The $m = 0$ negative energy, on the other hand, will gravitationally attract itself. Even if it is unclear whether the two energy forms attract or repel each other, it is not necessary to know this to solve for the evolutionary equations for this universe. As this universe expands, the prominence of the attractive $m = 0$ energy increases until it matches the energy density the neutral $n = 2$ energy.

A perhaps less-intuitive example is when positive $n = 3$ energy shares the universe with (only) negative $m = 2$. Here the $n = 3$ positive energy gravitationally attracts itself, while the $m = 2$ negative energy is gravitationally neutral toward itself. As this universe expands, the prominence of the neutral $m = 2$ energy increases until it matches the energy density of the attractive $n = 3$ energy. One might consider that the internal attractiveness of the universe decreases as it expands and the role of the gravitationally neutral negative energy becomes more important. Nevertheless, the deceleration of the universe might be considered as already set and sufficient to cause a turnover from expansion to contraction. An analogy is to consider the classic case of a $n = 3$ ball thrown into the air on the surface of the Earth. The ball loses kinetic energy, analogous here to ${\dot a}$, as it rises, but its deceleration is sufficient to cause it to ``turnover" at the top of its orbit and return to Earth.

%\subsection{Changes Between Stable Forms of Energy}

As discussed in detail in Paper I, it is possible for stable forms of energy to change into each other, and this applies to stable negative forms as well as stable positive forms. Rapid changes could occur---for example, as in the decay of an $n = 3$ form into an $n = 4$ forms: nuclear beta decay---among stable positive energy forms\cite{bahcall61}. An example of a slow change between stable negative energy forms would be the redshifting of negative radiation as it moved through the universe, where is slowly changes from $n = 4$ energy to $n = 3$ energy.

% Nifty summary Table 1 here.

\section{Summary: The past and future of the universe with stable negative energy forms}

We know that our present universe is composed predominantly of positive energy and is expanding\cite{blumenthal84,bucher99,huterer01,perlmutter98,zlatev99,spergel03}. Would it be possible for some stable negative energy form of negligible cosmological density in the present epoch but could grow to dominate and determine the future of the universe? The only way this could happen would be if some stable negative energy form with lower $n$ and $w$ than the current dominant stable energy form.  Since the dominant energy form is currently thought to be $n = 0$ ($w = -1$) dark energy, only a form of negative phantom energy could fill such a role.

Were any amount of stable negative phantom energy to exist in our universe, it would grow as the universe expands and eventually begin to dominate the universe. Since negative phantom energy is attractive such a component could stop the expansion of the universe and propel it into a contraction.  Even though the effect of this negative phantom energy would diminish and eventually become negligible, the universe would continue to contract, possibly ending in a big crunch.  Since this hypothetical stable negative phantom energy is undetectably small at present, it is really not possible to know with certainty that it does not exist, and therefore it is impossible to know the true future of our universe for this scenario.

Alternatively, would it be possible for some stable negative energy form of negligible cosmological density now but may have been significant in past of our observable universe? One  realization of this scenario would involve some stable negative energy form with higher $n$ and $w$ than the current stable positive energy form.  We know that our universe has some positive radiation and in at least two early epochs in the history of the universe, positive radiation was the cosmologically dominant energy form. 
Another realization demands a presence of stable negative ultralight in the current epoch.
% Such negative ultralight would have $n > 4$ and so increase in relative cosmological density as earlier times in the universe are considered. Such 
Negative ultralight energy would be gravitationally repulsive. Therefore it is possible that a previously contracting universe contained such negative ultralight, and this negative ultralight stopped the contraction and started an expansion.  Once the universe started expanding, this negative ultralight would again dissipate and again dilute to undetectable density. Therefore, since it is  not possible to confirm the existence of negative ultralight, it is not possible to know the true past of our universe for this scenario -- perhaps it started in a big bang or perhaps it underwent a big bounce.

Our universe could be an oscillating universe were it to contain even minuscule and undetectable amounts of both negative phantom energy and negative ultralight energy. As  $a$ becomes large, the negative phantom energy could suddenly begin to dominate and cause the universe to turnover and start contracting. This phantom energy would then again become effectively undetectable. As the universe contracts and its scale factor $a$ becomes small enough, the small amount of negative ultralight energy would suddenly dominate universe and, being repulsive, cause the universe to bounce. As the universe expands once again, the density of ultralight would again drop to undetectability and the cycle could repeat. Currently, it seems this type of universe cannot be ruled out theoretically\cite{steinhardt03}.

\begin{acknowledgements}
We thank the anonymous referee for valuable suggestions that have helped improve this paper.BP appreciates the help and support from Julian Varghese and Sandeep Ramachandran during the time of writing this paper.
\end{acknowledgements}

\nocite{*}
\bibliography{negativev2}

\end{document}